\documentstyle{pasj00}
\begin{document}

\SetRunningHead{A.Imada et al.}{SU UMa star 2QZ J021927.9-304545}

\title{The 2005 July Superoutburst of the Dwarf Nova 2QZ J021927.9-304545:
the SU UMa nature confirmed}

\author{Akira \textsc{Imada}, Taichi \textsc{Kato}}

\affil{$^1$Department of Astronomy, Faculty of Science, Kyoto University,
       Sakyo-ku, Kyoto 606-8502}
\email{a\_imada@kusastro.kyoto-u.ac.jp}

\author{L.A.G. \textsc{Monard}}

\affil{Bronberg Observatory, CBA Pretoria, PO Box 11426, Tiegerpoort
       0056, South Africa}

\author{Alon \textsc{Retter}}

\affil{Department of Astronomy $\&$ Astrophysics, Penn State 
       University, \\ 525 Davey Lab, University Park, PA 16802-6305, USA}

\author{Alex \textsc{Liu}}

\affil{Norcape Observatory, PO Box 300, Exmouth WA 6707, Australia}

\author{Daisaku \textsc{Nogami}}

\affil{Hida Observatory, Kyoto University, Kamitakara, Gifu 506-1314}


\maketitle

\begin{abstract}
We report on time-resolved photometry of the 2005 July superoutburst of
 the dwarf nova, 2QZ J021927.9-304545. The resultant light curves showed
 conspicuous superhumps with a period of 0.081113(19) days, confirming
 the SU UMa nature of the object. Although we missed the maximum
 phase of the outburst, the amplitude of the superoutburst well exceeded
 5 mag. This value is slightly larger than that of typical SU UMa-type
 dwarf novae. The superhump period decreased as time elapsed, as can be
 seen in most SU UMa-type dwarf novae. Based on the archive of ASAS-3,
 the recurrence time of a superoutburst of the variable turned out to be
 about 400 days. This value is typical of well known SU UMa stars. The
 distance to this system was roughly estimated as 370(+20, -60) pc using
 an empirical relation.
\end{abstract}

\section{Introduction}

Cataclysmic variables are close binany systems that consist of a white
dwarf primary and a late-type secondary star. The secondary star fills
its Roche-lobe, leading to mass transfer via the inner Lagragian point
($L_{1}$), forming an accretion disk around the white dwarf (for a review,
see e.g., \cite{war95book}; \cite{hel01book}). Dwarf novae are a
subclass of cataclysmic variables, further subdivided into three types:
U Gem stars, Z Cam stars, and SU UMa stars, based on their
outburst properties (for a review, \cite{osa96review};
\cite{kat04vsnet}). SU UMa-type dwarf novae, which have orbital periods
below 2 hours in most cases, exhibit two types of outburst: normal
outbursts whose duration is a few days and superoutbursts which continue
for about 2 weeks. Superoutbursts are always accompanied by ${\sim}$ 0.2
mag modulations called superhump. It is believed that a tidally deformed
eccentric accretion disk gives rise to phase-dependent tidal
dissipation, which is observed as superhumps (\cite{whi88tidal};
\cite{osa89suuma}). Optical photometry for SU UMa-type dwarf novae
during superoutbursts is one of the best ways to decipher the dynamics
of the accretion disk.

2QZ J021927.9-304545 is catalogued in 2dF QSO Redshift Survey (2QZ)
\citep{boy002qz1}. A
possible 2MASS counterpart has $J$ = 16.391(0.118), $H$ =
15.671(0.142), and $K$ = 15.359(0.202), respectively (for 2MASS
compilation of dwarf novae, see \cite{hoa02CV2MASS};
\cite{ima05j0137}). The object is identified with USNO B1.0 0592-0024305
($B$1 = 18.72, $R$1 = 18.49, $R$2 = 17.62, $I$2 = 18.02). Long-term
photometric monitoring of this
object was performed by the All Sky Automated Survey (ASAS,
\cite{poj02asas3}). ASAS-3 monitoring over the past few
years has provided some evidence that the star is a possible candidate
for an SU UMa-type dwarf nova. To test the SU UMa nature of
the system, we have continuously monitored the object at CBA Pretoria
since 2005 January. As a consequence, we detected an outburst on 2005
July 2 reaching a magnitude of about 11.9 in $R_{\rm c}$. However, bad
weather prevented us from time-resolved observations during the first
day of discovery. As for the ASAS-3 survey, the variable was invisible
on 2005 June 22 with the magnitude below the detection limit of the
ASAS-3, the object got brightened up to V ${\sim}$ 12.5 mag on 2005 July
3. After that we performed time-resolved CCD observations. Conspicuous
feature of superhumps was detected, so that we confirmed the SU UMa
nature of the object. Quiescent photometry of the object was carried out
at CBA Pretoria, yielding 18.2 in $R_{\rm c}$ (see e.g., [vsnet-alert
8521])\footnote{http://ooruri.kusastro.kyoto-u.ac.jp/pipermail/vsnet-alert/2005-July/000140.html}.

In the paper, we report on photometric results during the 2005 July
superoutburst of 2QZ J021927.9-304545, as well as long-term observations
of the object (hereafter aliased 2QZ 0219).

\section{Observations}

Time-resolved CCD observations were carried out at two sites of the
VSNET Collaboration team (see \cite{kat04vsnet}). A journal of
observations is summarized in table 1. Although we did not use any
filters, the obtained data were close to those of $R_{\rm
c}$$-$system. The exposure times were 15 and 30 sec, with a read-out
time of a few seconds. All frames were processed using AIP4WIN, and
magnitudes were derived by aperture photometry. As the comparison star, we
used GSC7007.1900, whose constancy was checked against several field
stars. The precision of individual data points was estimated to be
better than 0.03 mag. The calibrated magnitude for each site was
adjusted to ASAS-3 data of the object. 

Heliocentric correction was made for each data set before the following
analysis.

\begin{table}
\caption{Log of Observations.}
\begin{center}
\begin{tabular}{ccccc}
\hline\hline
HJD(start)$^*$ & HJD(end) & N$^\dagger$ & Exp(s)$^\ddagger$ &
 Observer$^\S$ \\
\hline
 3555.5776 & 3555.6681 & 498 & 15 & BM \\
 3556.5206 & 3556.6717 & 796 & 15 & BM \\
 3557.5063 & 3557.6535 & 808 & 15 & BM \\
 3558.5113 & 3558.6767 & 910 & 15 & BM \\
 3559.5070 & 3559.6761 & 932 & 15 & BM \\
 3561.5359 & 3561.6708 & 381 & 30 & BM \\
 3562.5185 & 3562.6683 & 380 & 30 & BM \\
 3563.4719 & 3563.6730 & 570 & 30 & BM \\
 3564.5204 & 3564.6707 & 426 & 30 & BM \\
 3565.3170 & 3565.4093 &  64 & 30 & AL \\
 3565.4987 & 3565.6409 & 394 & 30 & BM \\
 3566.5356 & 3566.6626 & 350 & 30 & BM \\
 3567.5472 & 3567.6694 & 345 & 30 & BM \\
\hline
\multicolumn{5}{l}{$^*$ HJD - 2450000. $^\dagger$ Number of frames.} \\
\multicolumn{5}{l}{$^\ddagger$ Exposure times.} \\
\multicolumn{5}{l}{$^\S$ BM: L.A.G. Monard, 32-cm telescope, SBIG ST-7XME} \\
\multicolumn{5}{l}{Pretoria, South Africa.} \\
\multicolumn{5}{l}{AL: Alex Liu, 30-cm telescope, SBIG ST-7e} \\
\multicolumn{5}{l}{Exmouth, Australia.} \\
\end{tabular}
\end{center}
\end{table}

\section{Results}

\subsection{light curve}

\begin{figure}
\begin{center}
\resizebox{80mm}{!}{\includegraphics{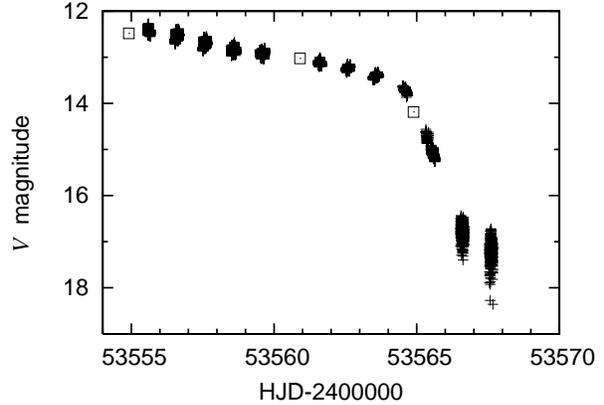}}
\end{center}
\caption{The entire light curve of the 2005 July superoutburst. The
 abscissa and the ordinate denote HJD$-$2400000 and $V$ magnitude,
 respectively. The open squares indicate the ASAS-3 photometry for
 which the typical error is within the size of the square. The duration
 of the plateau stage is at least 10 days, during which the object gradually
 fades at a rate of about 0.12 mag d$^{-1}$. }
\end{figure}

The overall light curve is represented in figure 1, in which the ASAS-3
observations are plotted by open squares as well. At onset of
our observation, 2QZ 0219 brightened up to 12.4 mag. After that,
the magnitude slightly declined at a rate of 0.12 mag d$^{-1}$, which is
a typical value of SU UMa stars. On HJD 2453565, 10 days after the epoch
of our first observation, the object plummeted at a rate of 1.54 mag d$^{-1}$,
after which 2QZ 0219 faded down to 17 mag. No rebrightening feature
was detected during our run. However, the presence of a rebrightening
after HJD 2453567 can not be ruled out.

\subsection{superhumps}

\begin{figure}
\begin{center}
\resizebox{80mm}{!}{\includegraphics{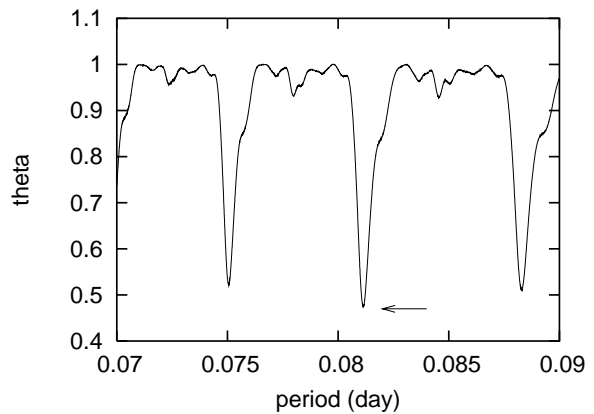}}
\end{center}
\caption{Theta diagram obtained by operating the PDM method to the data
 during the plateau stage after prewhitening. The best estimated period
 of superhumps is 0.081179(7) days.}
\end{figure}

Conspicuous superhumps were detected during the whole run. After
subtracting the linear declining trend, we performed a period analysis
using the phase dispersion minimization method (PDM)
\citep{ste78pdm}. We determined 0.081179 days as the best estimated
period of the superhump. Figure 2 shows the theta diagram for the plateau
stage. The error of the period was estimated using the Lafler-Kinman
class of methods, as applied by \citet{fer89error}.

\begin{figure}
\begin{center}
\resizebox{80mm}{!}{\includegraphics{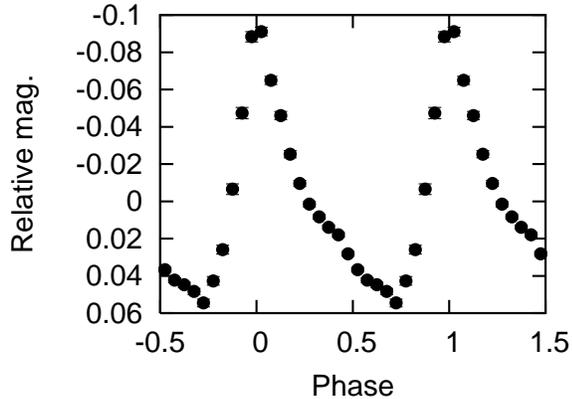}}
\end{center}
\caption{Phase-averaged superhumps during the plateau stage foleded by
 0.081179 days. A rapid rise and slow decline are characteristic of
 superhumps.}
\end{figure}

Figure 3 shows a phase-averaged superhump profile during the plateau
stage folded by 0.081179(7) days after subtracting a nightly decline
trend. As can be seen in figure 3, a rapid-rise and slow-decline, typical of superhump
profile among SU UMa stars, are remarkable. Daily-averaged superhump
profiles, folded by 0.081179(7) days, are depicted in figure 4. The shape
of the superhump variation was almost maintained with an amplitude of
${\sim}$0.2 mag except the last two days of our run.

\begin{figure*}
\begin{center}
\resizebox{53mm}{!}{\includegraphics{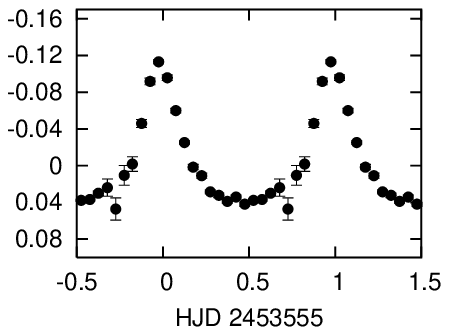}}
\resizebox{53mm}{!}{\includegraphics{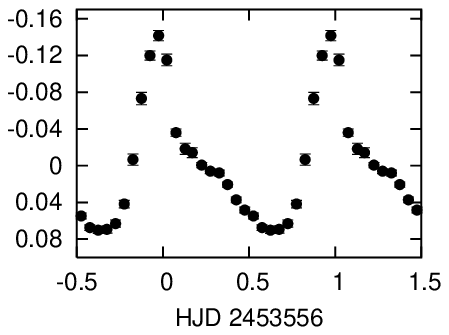}}
\resizebox{53mm}{!}{\includegraphics{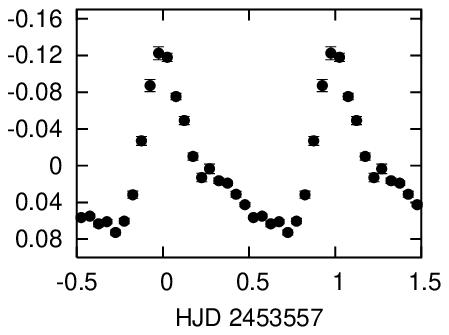}}
\resizebox{53mm}{!}{\includegraphics{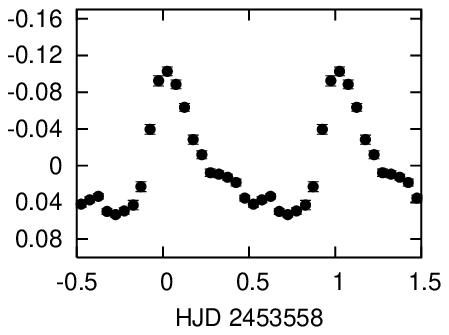}}
\resizebox{53mm}{!}{\includegraphics{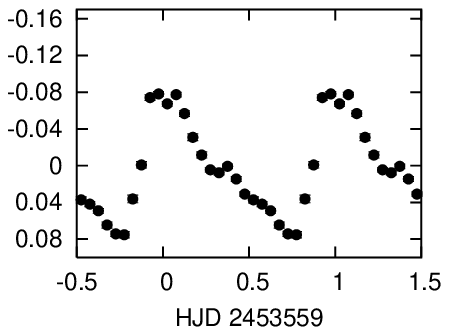}}
\resizebox{53mm}{!}{\includegraphics{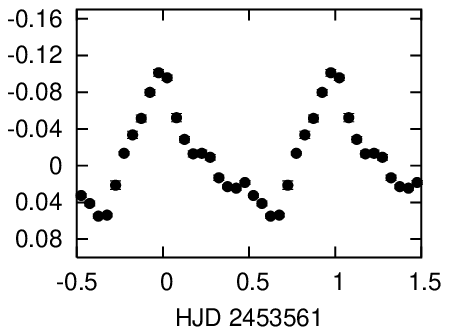}}
\resizebox{53mm}{!}{\includegraphics{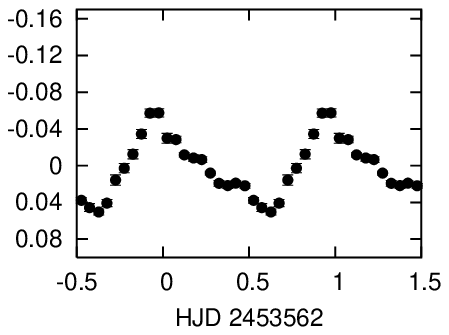}}
\resizebox{53mm}{!}{\includegraphics{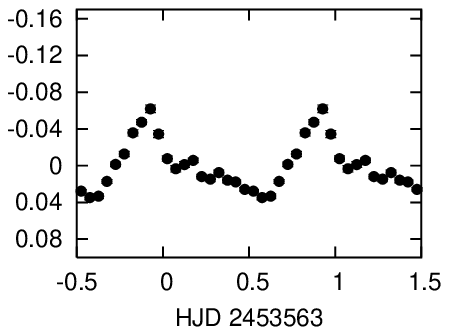}}
\resizebox{53mm}{!}{\includegraphics{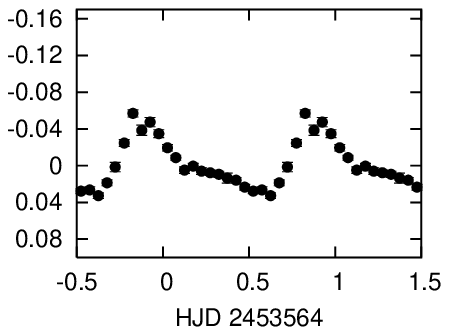}}
\resizebox{53mm}{!}{\includegraphics{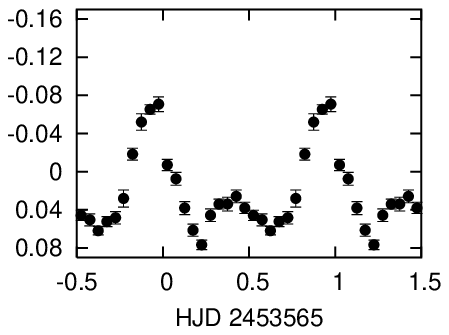}}
\resizebox{53mm}{!}{\includegraphics{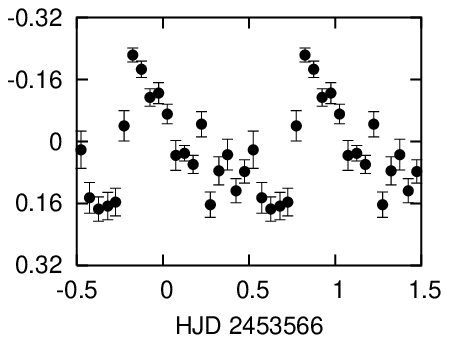}}
\resizebox{53mm}{!}{\includegraphics{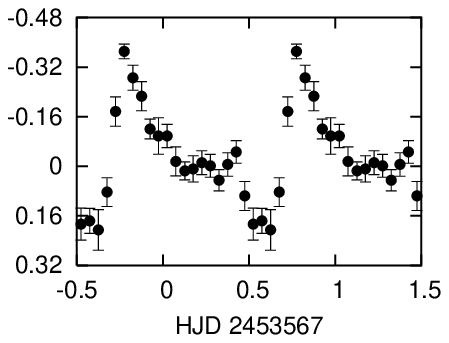}}
\end{center}
\caption{Daily averaged profile of the data folded by the 0.081179-d
 superhump period after subtracting the linear decline trend. The vertical
 and the horizontal axis denote the relative magnitude and phase,
 respectively. The epoch of phase is set on HJD 2455355.6025. The
 vertical axis is different for the last 2 runs.}
\end{figure*}

\subsection{superhump period change}

\begin{table}
\caption{Timing of superhump maxima.}
\begin{center}
\begin{tabular}{ccc}
\hline\hline
E$^*$ & HJD$^{\dagger}$ & err$^{\ddagger}$ \\
\hline
0 & 3555.5997 & 0.001 \\
12 & 3556.5737 & 0.003 \\
13 & 3556.6534 & 0.001 \\
24 & 3557.5484 & 0.001 \\
25 & 3557.6333 & 0.001 \\
36 & 3558.5258 & 0.002 \\
37 & 3558.6071 & 0.004 \\
49 & 3559.5766 & 0.001 \\
50 & 3559.6569 & 0.001 \\
74 & 3561.6093 & 0.001 \\
86 & 3562.5787 & 0.001 \\
87 & 3562.6615 & 0.001 \\
98 & 3563.5524 & 0.003 \\
99 & 3563.6333 & 0.002 \\
111 & 3564.6055 & 0.005 \\
123 & 3565.5734 & 0.002 \\
\hline
\multicolumn{3}{l}{$^*$ Cycle count.} \\
\multicolumn{3}{l}{$^{\dagger}$ HJD-2450000} \\
\multicolumn{3}{l}{$^{\ddagger}$ In the unit of day.} \\
\end{tabular}
\end{center}
\end{table}

We further examined the timing of the superhump maxima for the whole
superoutburst. The estimated superhump maxima are listed in table
2. The typical error of each maximum is of the order of 0.002 days. A linear
regression for the values listed in table 2 yielded the following
equation, 

\begin{eqnarray}
HJD(max) &= 0.081113(19) \times E + 53555.6025(13),
\end{eqnarray}

where the values in the parentheses denote the errors. By fitting the
deviation of the observed timings from the above calculation, we derived
the best fitted quadratic equation as follows:

\begin{eqnarray}
O - C =& -3.98(1.35)\times10^{-3} + 2.15(0.55)\times10^{-4} E \nonumber \\
       & -1.77(0.44)\times 10^{-6} E^{2}.
\end{eqnarray}

This equation indicates $P_{\rm dot}$ = $\dot{P}$/$P$ =
$-$4.4(1.1)$\times$10$^{-5}$, which is an ordinary value for SU UMa
stars (\cite{uem05tvcrv}; \cite{ole05v660her}). Figure 5 shows the
obtained $O - C$ diagram, as well as the best fitted quadratic curve. 

\begin{figure}
\begin{center}
\resizebox{80mm}{!}{\includegraphics{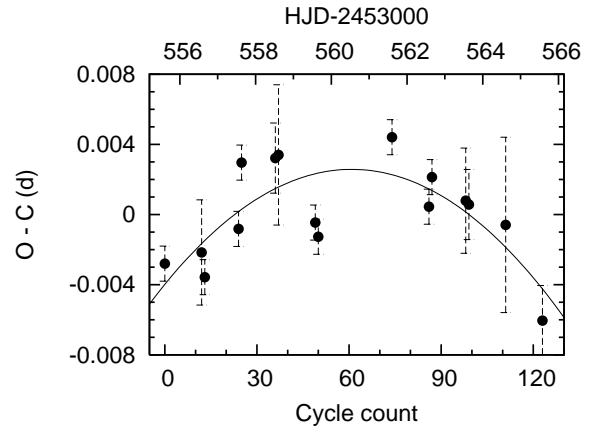}}
\end{center}
\caption{$O - C$ diagram of the superhump maximum timings. The abscissa
 and ordinate mean the cycle count and $O - C$, respectively. A linear
 regression ($C$) and the best quadratic fit are given in equation (1)
 and (2), respectively. Note that the superhump period decreased with
 time.}
\end{figure}

\section{Discussion}

\subsection{recorded outbursts}

\begin{figure*}
\begin{center}
\resizebox{160mm}{!}{\includegraphics{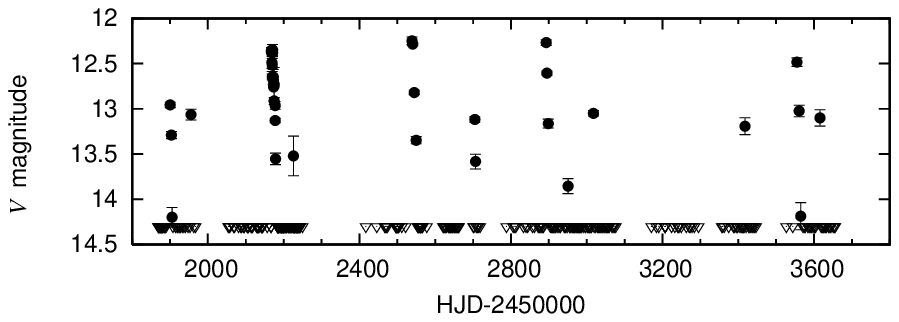}}
\end{center}
\caption{The long-term light curve of 2QZ 0219 taken from
the ASAS-3. The horizontal axis shows HJD-2450000, and the vertical axis
 shows $V$ magnitude. Filled circles and bottom triangles mean positive
 and negative observations, respectively. $V$ = 14.3 is the limiting
 magnitude.}
\end{figure*}

Thanks to the ASAS-3 survey \citep{poj02asas3}, we
can investigate the recorded outbursts over the past few
years. The long-term photometric behavior of 2QZ 0219 by ASAS-3 is
demonstrated in figure 6. From these data we judged the outburst type for
each outburst, and summarized it in table 3. Suspected superoutbursts
except the present one have been recorded on 2001 Sep. 15, 2002 Sep. 21,
and 2003 Sep. 10, so that we can roughly estimate the supercycle of the
object as follows:

\begin{equation}
T_{rec} = (3554.91624 - 2167.74163) / (3 + N),
\end{equation}

where N denotes the number of overlooked superoutbursts. With a little
algebra, we determine $T_{rec}$ = 346 $-$ 462 days. The
obtained supercycle length is a typical value of SU UMa-type
dwarf novae (\cite{kat03hodel}; \cite{nog97sxlmi}).

\begin{table}
\caption{Recorded outbursts}
\begin{center}
\begin{tabular}{cccccc}
\hline\hline
HJD & Mag & $T_{\rm min}$ & $T_{\rm max}$ & N & Type \\
(1) & (2) & (3) & (4) & (5) & (6) \\
\hline
1900.54153 & 12.96 & 5 & 18 & 3 & S: \\
1955.53592 & 13.06 & 1 & 10 & 1 & N \\
2167.74163 & 12.37 & 10 & 17 & 7 & S \\
2225.62825 & 13.52 & 1 & 9 & 1 & N \\
2538.78054 & 12.25 & 11 & 32 & 4 & S \\
2704.50660 & 13.12 & 2 & 6 & 2 & N \\
2892.83091 & 12.27 & 6 & 18 & 3 & S \\
2950.66662 & 13.86 & 1 & 6 & 1 & N \\
3017.66734 & 13.05 & 1 & 7 & 1 & N \\
3417.57925 & 13.19 & 1 & 5 & 1 & N \\
3554.91624 & 12.48 & 10 & 27 & 3 & S$^*$ \\
3615.88048 & 13.10 & 1 & 15 & 1 & N \\
\hline
\multicolumn{6}{l}{Columns. $-$ (1) HJD$-$2450000; (2) Maximum magnitude} \\
\multicolumn{6}{l}{detected by ASAS-3; (3) Minimum outbursting duration} \\
\multicolumn{6}{l}{in the unit of day; (4) Maximum outbursting duration} \\
\multicolumn{6}{l}{in the unit of day; (5) Number of frames; (6) Type} \\
\multicolumn{6}{l}{of outbursts. N: normal outburst; S: superoutburst} \\
\multicolumn{6}{l}{$^*$ This work.} \\
\end{tabular}
\end{center}
\end{table}

\subsection{distance}

If the orbital period of the system and the magnitude of its normal
outburst are known, one can roughly estimate the distance to the object
by an empirical relation for dwarf novae with a low-mid
inclination. According to \citet{war87CVabsmag}, the absolute magnitude of
dwarf novae is given as a function of its orbital period,

\begin{eqnarray}
M_{V} &= 5.64(13) - 0.259(24) \times P, 
\end{eqnarray}

where $M_V$ and P mean the absolute magnitude of dwarf novae in $V$ at
the maximum of the normal outburst, and the orbital period in the unit
of hours, respectively. Unfortunately, we have no information about the
orbital period of 2QZ 0219. Instead, we settle for using the obtained
superhump period as $P$. This prescription is fairly valid because there
is enough evidence that the difference between the superhump period and
orbital period is only a few percent. In addition, the absence of an
eclipse during the whole superoutburst indicate that 2QZ 0219 is not a
high inclination system. Hence, we can safely use the above equation in
order to estimate a distance to the object. With a little algebra, a
plausible distance to 2QZ 0219 lies in 370(+20, -60) pc. This distance
should be checked by other methods in future.

\subsection{superhump period change}
High speed CCD photometry has revealed that SU UMa-type dwarf novae
show changes of the superhump period during the superoutburst. Most SU
UMa-type dwarf novae show a decrease in their superhump periods as the
superoutburst proceeds, presumably due to the shrinkage of the disk radius,
or to a natural consequence of mass depletion from the disk
\citep{osa85SHexcess}. Recently, it has turned out that several SU
UMa-type dwarf novae including WZ Sge-type stars increase the superhump
period with time\footnote{Recently, V1028 Cyg \citep{bab00v1028cyg} and
TT Boo \citep{ole04ttboo} showed
both increasing and decreasing superhump period. Similar behavior was
also observed in V1974 Cyg \citep{ret97v1974cygSH}, although the object is
classified as a permanent superhump system .}. Figure 7 represents period $-$
$P_{\rm dot}$ diagram for SU UMa-type stars. As for the cause of increasing
superhump period, \citet{uem05tvcrv} has given an insightful suggestion
that the radius of the accretion disk at the outburst maximum is related
to an increase or decrease of the superhump period. If the accretion
disk is spread out well beyond the 3:1 resonance radius, where an
eccentric mode sets in, the eccentric mode can propagate beyond the 3:1
resonance radius because there exists plenty mass in the outer region of
the accretion disk during the superoutburst\footnote{WZ Sge stars and
large-amplitude SU UMa stars meet this condition.}. If the argument by
\citet{uem05tvcrv} is correct, we can naturally explain why the
increasing period of superhumps is ${\it only}$ seen among WZ Sge-type
stars and SU UMa stars with large amplitudes.

In the case of 2QZ 0219, taking into account the results of negative
$P_{\rm dot}$ derivative and the above mentioned suggestion by
\citet{uem05tvcrv}, the radius of the accretion disk was not so large at
the superoutburst maximum, at most, as much as the 3:1 resonance
radius. Thus, the superhump period variation for 2QZ 0219 behaves as the
``textbook'' SU UMa-type dwarf novae. In order to test the validity of
Uemura's implication, hydrodynamic simulation should be performed.

\begin{figure}
\begin{center}
\resizebox{80mm}{!}{\includegraphics{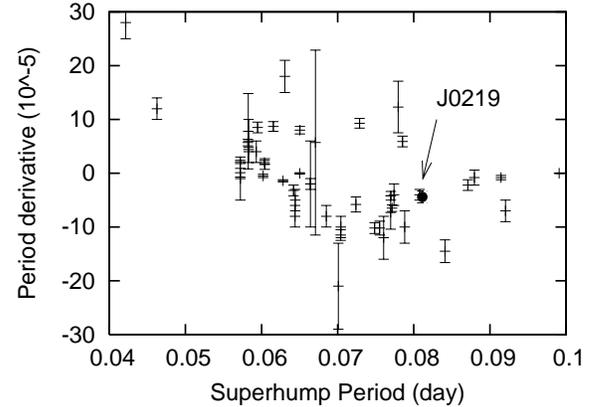}}
\end{center}
\caption{$P_{\rm sh}$$-$$\dot{P}$ diagram of $O - C$ variation explored SU
 UMa stars. The location of 2QZ 0219 is designated as a
 filled circle. The data were originally taken from \citet{ima05gocom}.}
\end{figure}

\section{Summary}

The present photometric observations of 2QZ 0219 allowed us to
confirm a new member of SU UMa-type dwarf novae by detecting
superhumps. The best estimated superhumps period of J0219 was
0.081113(19) days = 116.8 min, placing 2QZ 0219 near the lower edge of the
period gap. In conjunction with our observations and the ASAS-3 archival
data, the plateau stage of the superoutburst lasted at
most 20 days. The amplitude of the superoutburst exceeded 5 mag,
slightly larger than ordinary SU UMa-type dwarf novae but within the
normal range. An estimation of the recurrence time of the superoutburst
yielded ${\sim}$ 346 $-$ 462 days. We can roughly determine the distance
to 2QZ 0219 to be 370(+20, -60) pc. The resultant $O - C$ diagram,
together with the above results, led us to the conclusion that 2QZ 0219
is a new member of a prototypical SU UMa-type dwarf nova.

\vskip 3mm

We are grateful to many VSNET observers who have reported vital
observations. We would also express our gratitude to G. Pojmanski for
providing invaluable data of ASAS-3 observations. This work is supported
by a Grant-in-Aid for the 21st Century COE ``Center for Diversity and
Universality in Physics'' from the Ministry of Education, Culture,
Sports, Science and Technology (MEXT). This work is partly supported by
a grant-in aid from the Ministry of Education, Culture, Sports,
Science and Technology (No. 16340057, 17740105).


\begin{thebibliography}{}

\bibitem[Baba et~al.(2000)]{bab00v1028cyg}
  Baba, H., Kato, T., Nogami, D., Hirata, R., Matsumoto, K., \& Sadakane, K.\
  2000, \pasj, 52, 429

\bibitem[Boyle et~al.(2000)]{boy002qz1}
  Boyle, B.~J., Shanks, T., Croom, S.~M., Smith, R.~J., Miller, L., Loaring,
  N., \& Heymans, C.\ 2000, \mnras, 317, 1014

\bibitem[Fernie(1989)]{fer89error}
  Fernie, J.~D.\ 1989, \pasp, 101, 225

\bibitem[Hellier(2001)]{hel01book}
  Hellier, C.\ 2001, Cataclysmic Variable Stars: how and why they vary
  (Berlin: Springer-Verlag)

\bibitem[Hoard et~al.(2002)]{hoa02CV2MASS}
  Hoard, D.~W., Wachter, S., Clark, L.~L., \& Bowers, T.~P.\ 2002, \apj, 565,
  511

\bibitem[Imada et~al.(2005)]{ima05gocom}
  Imada, A., {et~al.}\ 2005, \pasj, 57, 193

\bibitem[Imada et~al.(2005)]{ima05j0137}
  Imada, A., {et~al.}\ 2005, astro-ph/0510764

\bibitem[Kato et~al.(2003)]{kat03hodel}
  Kato, T., Nogami, D., Moilanen, M., \& Yamaoka, H.\ 2003, \pasj, 55, 989

\bibitem[Kato et~al.(2004)]{kat04vsnet}
  Kato, T., Uemura, M., Ishioka, R., Nogami, D., Kunjaya, C., Baba, H., \&
  Yamaoka, H.\ 2004, \pasj, 56S, 1

\bibitem[Nogami et~al.(1997)]{nog97sxlmi}
  Nogami, D., Masuda, S., \& Kato, T.\ 1997, \pasp, 109, 1114

\bibitem[Olech et~al.(2004)]{ole04ttboo}
  Olech, A., Cook, L.~M., Z{\l}oczewski, K., Mularczyk, K., K\c{e}dzierski, P.,
  Udalski, A., \& Wisniewski, M.\ 2004, Acta Astron., 54, 2330

\bibitem[Olech et~al.(2005)]{ole05v660her}
  Olech, A., Z{\l}oczewski, K., Cook, L.~M., Mularczyk, K., K\c{e}dzierski, P.,
  \& Wisniewski, M.\ 2005, Acta Astron., 55, 237

\bibitem[Osaki(1985)]{osa85SHexcess}
  Osaki, Y.\ 1985, \aap, 144, 369

\bibitem[Osaki(1989)]{osa89suuma}
  Osaki, Y.\ 1989, \pasj, 41, 1005

\bibitem[Osaki(1996)]{osa96review}
  Osaki, Y.\ 1996, \pasp, 108, 39

\bibitem[Pojmanski(2002)]{poj02asas3}
  Pojmanski, G.\ 2002, Acta Astron., 52, 397

\bibitem[Retter et~al.(1997)]{ret97v1974cygSH}
  Retter, A., Leibowitz, E.~M., \& Ofek, E.~O.\ 1997, \mnras, 286, 745

\bibitem[Stellingwerf(1978)]{ste78pdm}
  Stellingwerf, R.~F.\ 1978, \apj, 224, 953

\bibitem[Uemura et~al.(2005)]{uem05tvcrv}
  Uemura, M., {et~al.}\ 2005, \aap, 432, 261

\bibitem[Warner(1987)]{war87CVabsmag}
  Warner, B.\ 1987, \mnras, 227, 23

\bibitem[Warner(1995)]{war95book}
  Warner, B.\ 1995, Cataclysmic Variable Stars (Cambridge: Cambridge
			    University Press)

\bibitem[Whitehurst(1988)]{whi88tidal}
  Whitehurst, R.\ 1988, \mnras, 232, 35

\end{thebibliography}
\end{document}